# Role of Oxygen Functionalities in Graphene Oxide Architectural Laminate Subnanometer Spacing and Water Transport

August 2016

C. A. Amadei[1], A. Montessori[2], J. P. Kadow[1,3], S. Succi[1,4], C. D. Vecitis[1,*]

[1]*John A. Paulson School of Engineering and Applied Sciences, Harvard University, Cambridge, MA 02138, USA*
[2]*Dept. of Engineering - University of Rome "Roma Tre", 00141 Rome – Italy*
[3]*Technische Universität München*
[4]*Istituto per le Applicazioni del Calcolo, CNR, 00185 Rome - Italy*

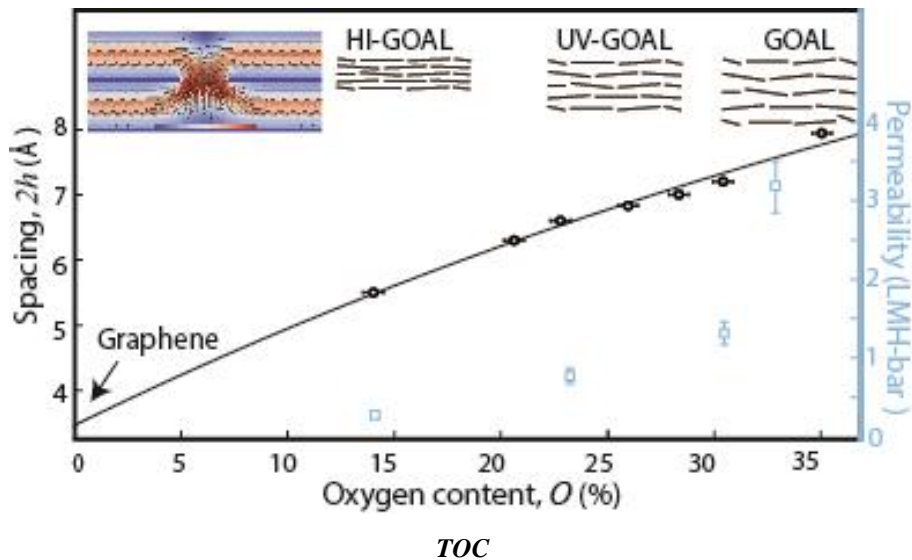

TOC

[*]*Corresponding author: Chad D. Vecitis,*

Email: *vecitis@seas.harvard.edu*, *Phone: (617) 496-1458*
Address: *Paulson School of Engineering and Applied Sciences, Harvard University, Cambridge, MA 02138*

**Abstract**

Active research in nanotechnology contemplates the use of nanomaterials for engineering applications. However, a primary challenge is understanding the effects of nanomaterial properties on industrial device performance and translating unique nanoscale properties to the macroscale. One emerging example is graphene oxide (GO) membranes for separation processes. Thus, here we investigate how individual GO properties can impact layered GO characteristics and water permeability. GO chemistry and morphology were controlled with easy-to-implement photo-reduction and sonication techniques and were quantitatively correlated offering a valuable tool to speed up the characterization process. For example, one could perform chemical analysis and concurrently obtain morphology information or vice versa. Chemical GO modification allows for fine control of GO oxidation state and GO laminate nanoarchitecture enabling controlled synthesis of a GO architectural laminate (GOAL). The GOAL can be considered as the selective layer of the membrane created by interconnected sub-nanometer channels, characterized by a length and a height (i.e., GO spacing), through which water molecules permeate. Water permeability was measured for eight GOAL characterized by different GO chemistry and morphology, and indicate that GO nanochannel height dictates water transport. For instance, a 10% decrease in the GO nanochannel height results in a >50% reduction in the water permeability. The experimental outputs were corroborated with mesoscale simulations of relatively large domains (>1000s nm$^2$). The simulations indicate a no-slip Darcy-like water transport regime inside the GOAL due to the presence of basal oxygen functionalities. The experimental and simulation evidence presented in this letter help create a clearer picture of water transport in GO and can be used to rationally design more effective and efficient GO membranes.

Membrane technology has climbed to a paramount role in engineering applications such as water treatment,[1,2] gas separation,[3] energy storage,[4] and food packaging.[5] Recent works demonstrate that graphene could represent a valid alternative to traditional polymeric membranes[6,7] due to its mechanical strength,[8] chemical stability,[9] near frictionless water transport,[10,11] and the potential to engineer well-defined surface pores and/or internal nanochannels.[12,13] Regarding the pore structure, graphene-based membranes can be classified into two main groups: nanoporous graphene (NPG) and graphene oxide laminates (GOL).[14] NPG is obtained by creating pores in a pristine graphene monolayer via ion-bombardment and/or etching.[15,16,17] GOL are synthesized by casting graphene oxide (GO) flakes on porous substrates. The flakes spontaneously arrange in an ordered fashion to form a laminate structure characterized by interconnected GO nanochannels.[18,19]

The study of water transport in elemental carbon nanomaterials began in the mid-2000s, when fast (slip) water transport within nanometer-wide 1D carbon nanotubes was reported.[20,21] In 2012, study of water transport in carbon materials expanded to 2D carbon nanoarchitectures (such as NPG and subsequently GOL).[22,23] GOL water transport has been modeled using molecular dynamics (MD), and results indicated a fast flow due to the low wall friction of water within GO nanochannels or to the presence of defects in the GO flakes.[24,25] Experimental studies also observed GOL fast water transport,[26,27] although a clear consensus on the mechanism (low wall friction versus nanometer defects) has not yet been reached.

In this letter, we provide insight into the mechanism of water flow within GOL via a combination of pressure-driven pure water permeability experiments and mesoscale Lattice-Boltzmann (LB) fluid dynamics simulations. Ultrathin graphene oxide architectural laminates (GOAL, Figure 1 left) were synthesized by vacuum filtration (VF) then subsequently photo-reduced by UV irradiation (in ambient or vacuum) or hydroiodic acid (HI) and the flake size was reduced by GO sonication prior to VF (Figure 1, right). The effect of GO surface chemistry and flake size on GOAL nanochannel architecture, e.g., interlayer spacing ($2h$) and tortuosity ($l$), was characterized by scanning electron microscopy (SEM), X-ray diffraction (XRD), and X-ray photoelectron spectroscopy (XPS). GOAL pure water permeability was determined in dead-end filtration mode and mesoscale LB simulations were used to elucidate hydrodynamic phenomena.

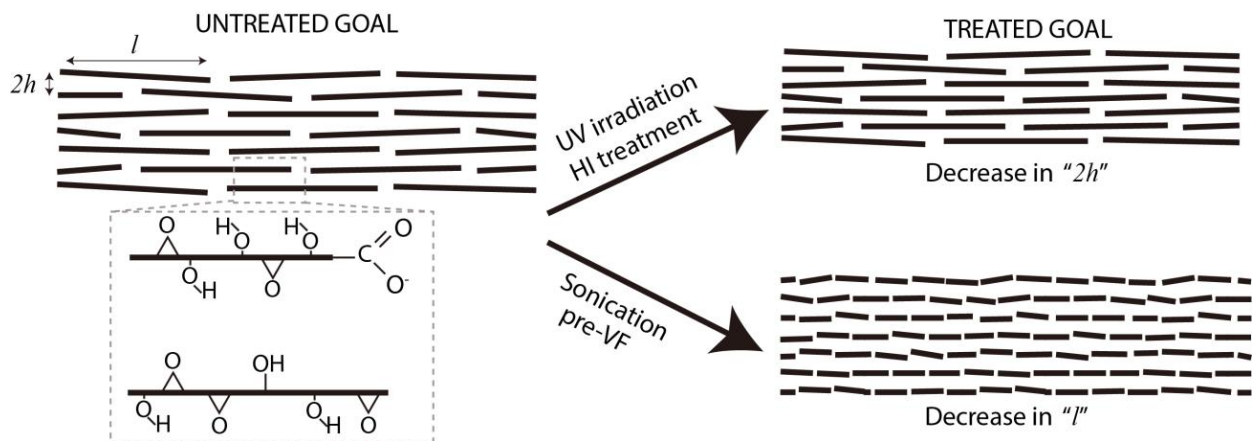

**Figure 1: Scheme representing the modification of GOAL.** The main two dimensions (2h and l) of the nanochannels' architecture are independently modified with easy-to-implement techniques and are drawn to scale.

GOAL are cast on polyvinylidene difluoride (PVDF) ultrafiltration membrane via VF and their morphology is determined by scanning electron microscopy (SEM, Figure 2). In Figure 2a right, the bare PVDF membrane is characterized by a flower-like structure with a pore size 310±126 nm. After VF, the GOAL homogenously coats the PVDF (Figure 2a, left), but the flower-like structure of the PVDF is still observed (inset in Figure 2a), due to the ultrathin (47±6 nm) GOAL layer (mass and SEM analysis). A cross-section image of the GOAL/PVDF highlights the ultrathin GOAL and the preservation of the underlying PVDF morphology. Morphology conservation is important for membranes since specific nano- and micro-morphology affects fouling.[28] GOAL morphology and thickness are also corroborated by Atomic Force Microscopy (AFM) images (Figure S1).

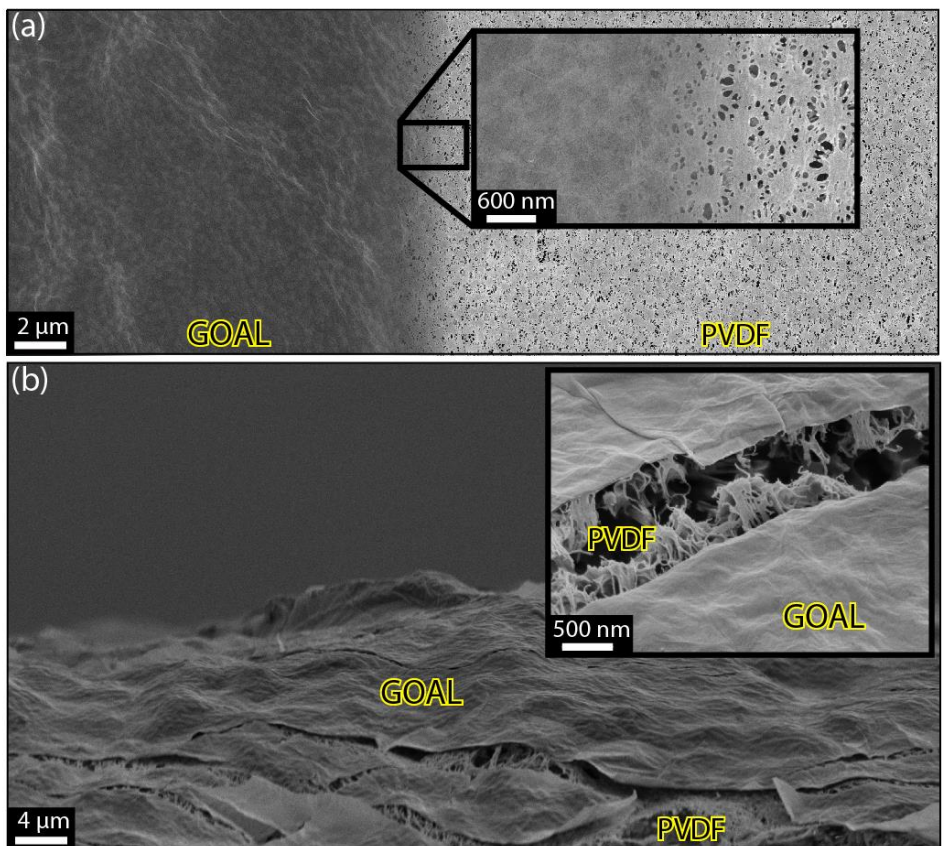

**Figure 2: GOAL morphology.** (**a**) Areal top-down and (**b**) cross-section images of GOAL membranes obtained via SEM.

The GOAL/PVDF are then subjected to reductive treatment, such as UV irradiation: 15 to 1440 min (Figure S2). Previous reports indicate that the effects of UV irradiation[29,30] on GO will depend on the environment (e.g., water or air) and can dictate whether GO reduction occurs.[31] In this work, we for the first time explore the UV reduction of GO in vacuum, which may reduce photochemical degradation and increase conductivity.[32] The evolution of the GOAL XPS C1s spectrum with UV irradiation time is displayed in Figure 3a. After 1440 min, GOAL graphitization is indicated by an increase of the carbon-carbon (C-C and C=C: binding energy ≈285 eV) bond peak percentage of total C-bonds from 28±2 to 71±2%. Correspondingly, the relative single carbon-oxygen bond peak intensity such as hydroxide (C–OH, ≈287 eV), decreases from 69±1 to 14±1%. The oxygen to carbon mass (i.e., O/C) ratio decreases from 72±2 to 35±2% from the untreated GOAL to the 1440 min-UV–irradiated GOAL, respectively. The XPS characterization data is summarized in Table S1 and Eq.1 represents a possible chemical reaction during UV irradiation where an epoxide is photolyzed to a C=C double bond:

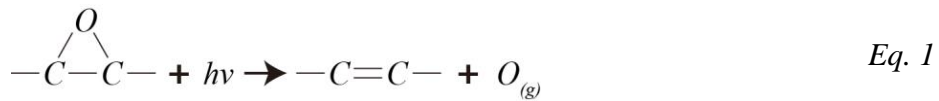

$$-\overset{\overset{O}{\diagup\diagdown}}{C}-C- + h\nu \rightarrow -C=C- + O_{(g)} \qquad Eq.\ 1$$

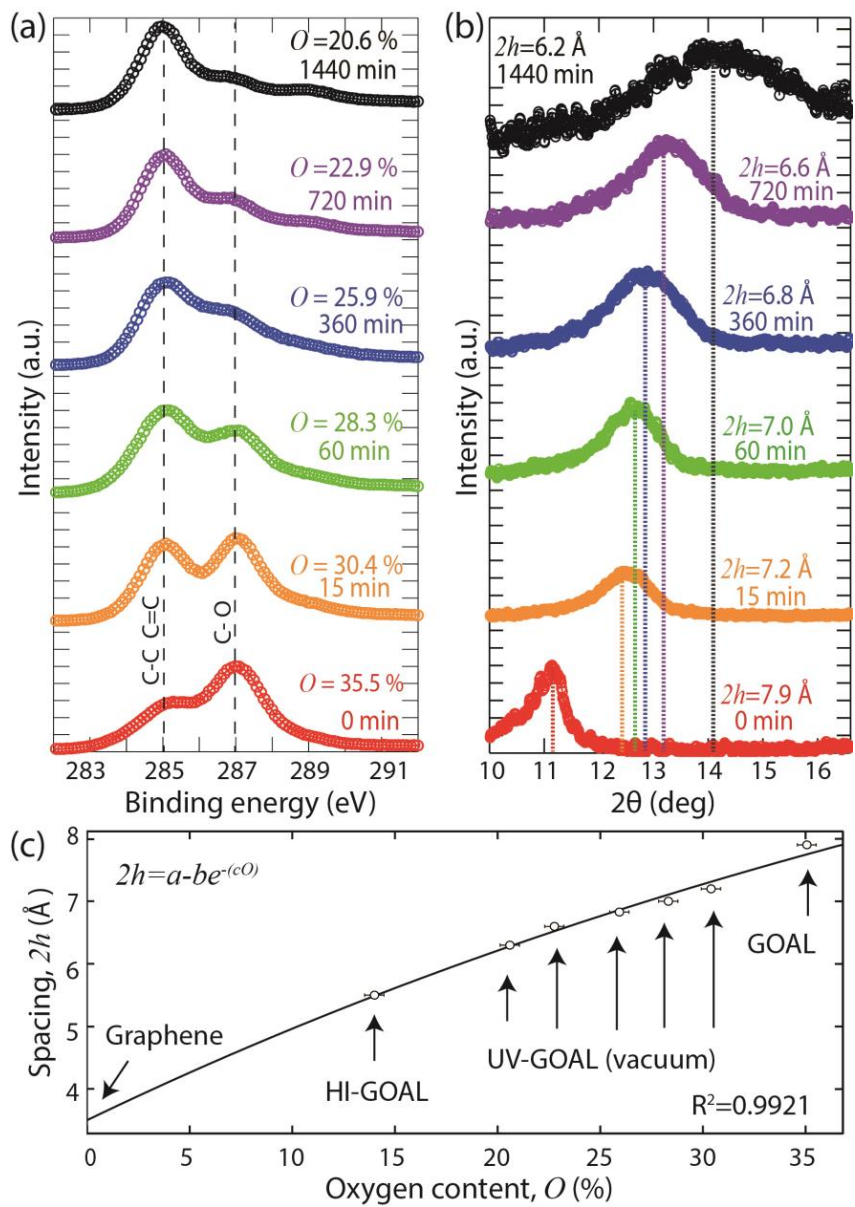

**Figure 3: GOAL interlayer spacing as a function of oxygen content.** (**a**) XPS and (**b**) XRD time-dependent spectra for UV irradiated GOAL membranes in vacuum. XRD FWHM peak is equal to 0.7, 1, 1.2, 1.3, 2, and 3.5° for 0, 15, 60, 360, 720, and 1440 min irradiation time, respectively. (**c**) GOAL interlayer spacing as a function of GO oxygen content can be fitted with an exponential rise to maximum GO interlayer spacing. In the equation, a, b, and c are empirical coefficients with values of 29.6, 26 and 0.004, respectively.

The GOAL XRD spectrum as a function of UV irradiation time is displayed in Figure 3b. All spectra display a peak corresponding to the GOAL interlayer spacing ($2h$) (Figure 1, left). The interlayer spacing is quantitatively correlated to the GOAL oxygen content indicating basal plane oxy-groups mediate GOAL interlayer spacing. Specifically, UV reduction of oxy-functional groups results in an XRD peak shift of 2.7 Å or ≈25% decrease in $2h$, 7.9 Å for untreated GOAL and 6.2 Å for GOAL after 1440 min UV treatment. Elemental carbon reduction has been observed to be environment dependent,[33] thus we compared the reduction via UV irradiation in vacuum versus ambient air. The GOAL O/C ratio after 60 min UV in vacuum is similar to 360 min in air indicating significantly faster reduction kinetics and the final O/C ratio after 1440 min is 0.35 and 0.46 for UV irradiation in vacuo and air, respectively (see Table S1). GOAL UV irradiation in air results in slower reduction likely due to ambient $O_2$ propagating reactions with GO surface radicals and slower removal of UV-produced gaseous radicals (Eq.1) similar to phenomena during thermal reduction of GO.[34,35] Correspondingly, UV GOAL treatment in air results in broader XRD peaks indicating a more disordered structure compared to UV treatment in vacuum. For example, the full width half maximum (FWHM) after 60 min UV is ≈1.2° and ≈1.8° for vacuum and ambient air, respectively (Figure S3b). Alternatively, the GOAL were also reduced by conc. HI at 50 °C for 1 min. HI reduces the GOAL to a greater extent than UV, in accordance with previous reports,[36,37] resulting in an O/C <0.15 and a interlayer spacing of 5.5 Å (Figure S4). Again, HI reduction of GOAL oxy-functional groups lead to a decrease in interlayer spacing, and supports a possible relationship between the two variables.

GOAL interlayer spacing ($2h$) as a function of oxygen content ($O$) is displayed in Figure 3c and well fit ($R^2 > 0.99$) by an exponential rise to a maximum GO interlayer spacing:

$$2h = a - b * \mathrm{Exp}(-c * O) \qquad \text{Eq. 2}$$

where $a$, $b$ and $c$ are empirical coefficients with values of 29.6, 26, and 0.004, respectively. When $O$ is zero (i.e., pristine graphene), Eq. 2 yields $2h = 3.63$ Å, similar to the theoretical spacing for pristine graphene (3.4 Å) supporting the theoretical reliability of the trend in Figure 3c. In the case of completely oxidized GO ($O$ equal 100, one O atom for one C atom), Eq. 2 yields $2h = 12.2$ Å, the upper limit for GO spacing. The small $c$ value indicates that large changes in $O$ will result in sub-angstrom variation in GOAL interlayer spacing. The relation in Eq. 2 has potential to significantly decrease GOL characterization time by performing a single characterization (e.g., XPS or XRD) to determine both oxygen content and interlayer spacing. Note that Eq. 2 currently applies to our synthesized GOAL, but deviations from this may be expected due to alternative casting techniques and/or varying GO properties (e.g., flake size and/or oxy-functional group content/distribution).

GOAL reduction results in a corresponding reduction of GO nanochannel spacing. However, it is also possible to rationally modify the length (*l*) of the nanochannels (Figure 1, right) by ultrasonic irradiation of the GO solution before VF casting.[38] For example, the GO flake area decreased by more than one order of magnitude, from 52.2 ±18.9 µm$^2$ to 1.3 ±0.4 µm$^2$ (Figure S5) after 23 min of bath sonication corresponding to a decrease in *l* by a factor of ≈7, assuming a square flake. Note that the modification of *2h* and *l* are not interdependent as it is possible to modify one dimension without affecting the other (Figure S6).

Now that we have demonstrated the ability to tune primary GOAL dimensions (*2h* and *l*) by exploiting GO metastability, the GOAL water permeability as a function of GO nanoscopic properties is investigated. Attention was focused on eight GOAL: i-ii) Untreated GOAL (0-GOAL and 0-GOAL-son), iii-iv) 15 min UV-irradiated GOAL in vacuum (15-GOAL and 15-GOAL-son), v-vi) 360 min UV-irradiated GOAL in vacuum (360-GOAL and 360-GOAL-son), and vii-viii) HI treated GOAL (HI-GOAL and HI-GOAL-son). Note that -son refers to membranes obtained by casting a GO solution which was sonicated, thus characterized by a smaller GO flake size. The pure water permeabilities were obtained via dead-end filtration (Figure S7) and are summarized in Figure 4.

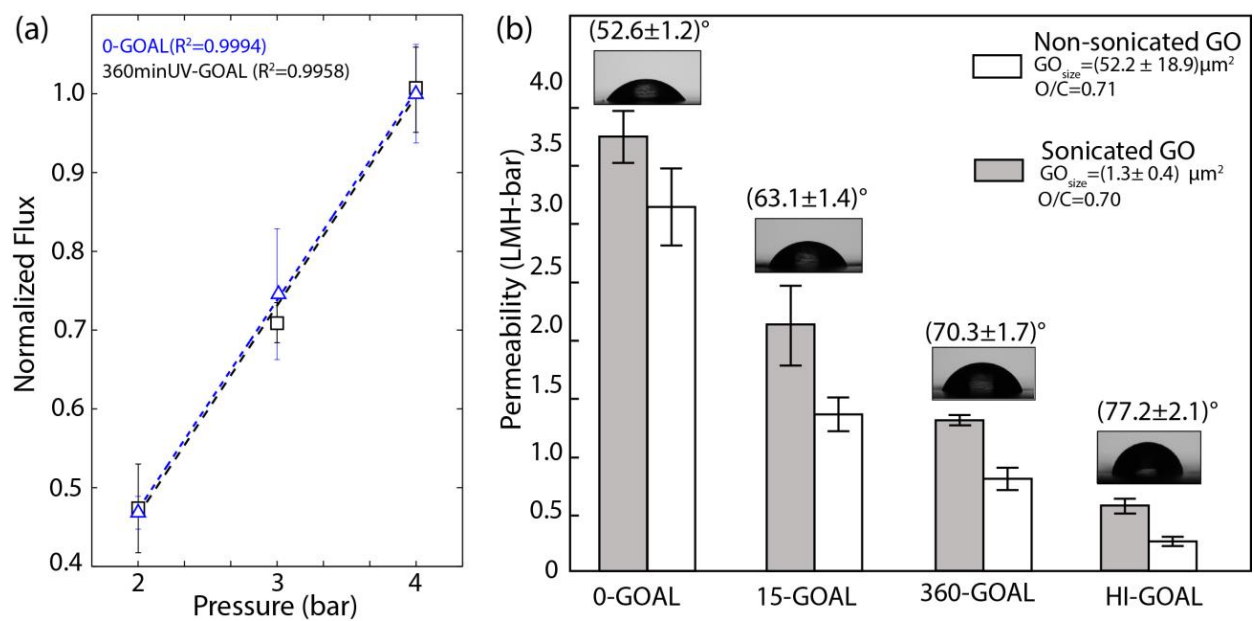

**Figure 4: GOAL Pure water permeability. (a)** Darcy behavior of the UV treated GOAL membranes in vacuum. **(b)** Permeability and SCA of eight GOAL membranes and variations of SCA. Each permeability was evaluated at least in duplicate

The normalized flux for two membranes (0-GOAL (blue) and 360-GOAL (black)) at three pressures (2, 3, and 4 bar) is presented in Figure 4a. The water flux is linearly dependent on the applied pressure indicating Darcy behavior. In the Darcy hydrodynamic regime, water transport follows the no-slip condition.[39,40] The permeability in LMH-bar (i.e., L m$^{-2}$ h$^{-1}$ bar$^{-1}$) for the eight GOAL membranes is displayed in Figure 4b. In terms of absolute values, the permeability varies from 3-4 LMH-bar to 0.25-0.75 LMH-bar, in accordance with values reported for reduced ultrathin GO membranes on polymer substrates.[26-28,41] Although higher values (>10s LMH-bar) of permeability have been reported for GO membranes, these values may be related to the presence of defects or larger GO nanochannels than those observed here due to the reduction of the GO membranes in air.[42,43] Increased permeability (>100 LMH-bar) can also be achieved by intercalating GO with high aspect ratio nanostructures such as rods.[13,44] Discrepancies between permeability values reported in Figure 4b and previous literature values could also emerge from the fact that previous studies did not evaluate the membrane performance under steady state conditions (always lower than initial permeability). For example, the time-dependent permeate volume (Figure S8) displays that the GOAL flow rate decreases with time until reaching steady state conditions at ≈700 min, similar to polymer membranes, the GOAL initially compresses after pressure is applied likely due to GO flake rearrangement.[45,44] For example, 0-GOAL have an initial permeability 3-4-fold greater than the steady-state permeability.

A clear decrease in GOAL permeability is also observed upon GO chemical reduction (Figure 4b). For example, a 6-7-fold reduction in permeability is observed for HI-GOAL (0.54 LMH-bar) compared to 0-GOAL (3.77 LMH-bar) even though there is only a 50% reduction in *2h*. UV irradiation time also leads to a decrease in GOAL permeability; 2.11 LMH-bar and 1.29 LMH-bar for 15-GOAL (2h=7.2 Å) and 360-GOAL (2h=6.8 Å), respectively. In summary, permeability is predominantly related to a decrease in GO interlayer nanochannel spacing, which in turn is related to the extent of GO chemical reduction, and there is little effect in reducing GO length by sonication.

The permeability decrease with reduction indicates that the process does not create macroscopic defects (i.e., holes) in the GOAL structure, which would result in an increased permeability,[43] and instead favors an increase of interlayer π-π GO interactions and reduction in nanochannel spacing (*2h*).[46] Increased π-π interaction is confirmed by Raman spectroscopy as the D/G peak intensity ratio ($I_D/I_G$) decreases with increasing UV irradiation time, confirming the restoration of the graphitic domains via GOAL reduction (Figure S9). $I_D/I_G$ is 1.1 for untreated GO,[38] and decreases to 0.95 and 0.90 for GO UV after for 15 min and 720 min irradiation in vacuum, respectively. GO reduction also increases in hydrophobicity with the static contact angle (SCA) increasing from 55.6 ±1.2º (0-GOAL) to 77.2 ±2.1º (HI-GOAL).[47,48] In particular, the increase in hydrophobicity is exponentially dependent with the extent of GO reduction (see Eq. S1 and Figure S10). The theoretical reliability of this trend is confirmed by the fact that Eq. S1 yields SCA=82.6 º for pristine graphene/graphite (O=0%), which is corroborated by several experimental and simulations reports.[49,50,51] The increase in hydrophobicity and may also affect GOAL permeability.

For comparison, sonication increased GOAL permeability by only 0.6±0.3-fold (grey bars; Figure 4b), even though *l* decreased by ≈7-fold. The permeability increase is related to the decrease in membrane tortuosity caused by a decrease in length (*l*) of the GO nanochannels. Considering HI-GOAL, it is of note that a *2h* decrease of 30% (7.9 to 5.5 Å) results in an ≈90% decrease in water permeability, in contrast a decrease of ≈700% in *l* (≈7 to ≈1 µm), only resulted in an increase of ≈100% in the water permeability suggesting a 2D regime that does not follow Hagen-Poiseuille flow.

The change in the permeability ($\Delta k$) was divided by the respective change in the dimension for the two scenarios ($\Delta 2h$ and $\Delta l$), in order to obtain normalized $\Delta k_l$ and $\Delta k_{2h}$, where subscripts indicate dimension in question. The $\Delta k_{2h}$ values range from 2.80-6.35, whereas $\Delta k_l$ values range from 0.02 to 0.1 (Table S2 and S3). The orders of magnitude difference between the two ranges highlights that the characteristic dimension dominating permeability is *2h*, whereas in comparison *l* has a limited effect on permeability.

In order to gain a deeper understanding of the GOAL water transport mechanism, a Lattice-Boltzmann (LB) approach augmented with a novel Langevin-like frictional forcing term between water molecules and the GO nanochannels basal plane oxy-functional groups via H-bonding was employed (Figure 5). The model geometry and the characteristic water flow path within the GO nanochannel is displayed in Figure 5a. Of note is the large simulation domain (1000s nm$^2$), which would be computationally costly for simulation tools currently used to evaluate water transport within GO nanochannels (e.g., MD).[25,52] The $\Delta k_{2h}$ versus GO nanochannel height variation ($\Delta 2h$) is plotted in Figure 5b. Simulations were carried out for four different interlayer spacings of 7.2, 6.7, 6.3, and 6.1 Å and for a flake size equal to the sonicated flake (≈1 µm). We observe that $\Delta k_{2h}$ decreases with a decrease in GO interlayer spacing from 5 (*2h*=7.2 Å) to 3 (*2h*=6.1 Å), in agreement with experimental range (2.80<$\Delta k_{2h}$<6.35). Thus, the simulations support the decreasing trend of the permeability with reduction of the GOAL interlayer spacing. Of note for both experimental and simulations outputs, the smallest variations in the spacing (<10%) lead to the largest normalized variations in the permeability (>5). This spacing variation could also be achieved by unintentional reduction of the GOAL (i.e., prolonged exposure to sunlight) and highlights how GO metastability[19,53] may affect the macroscopic performance of GO-based applications.

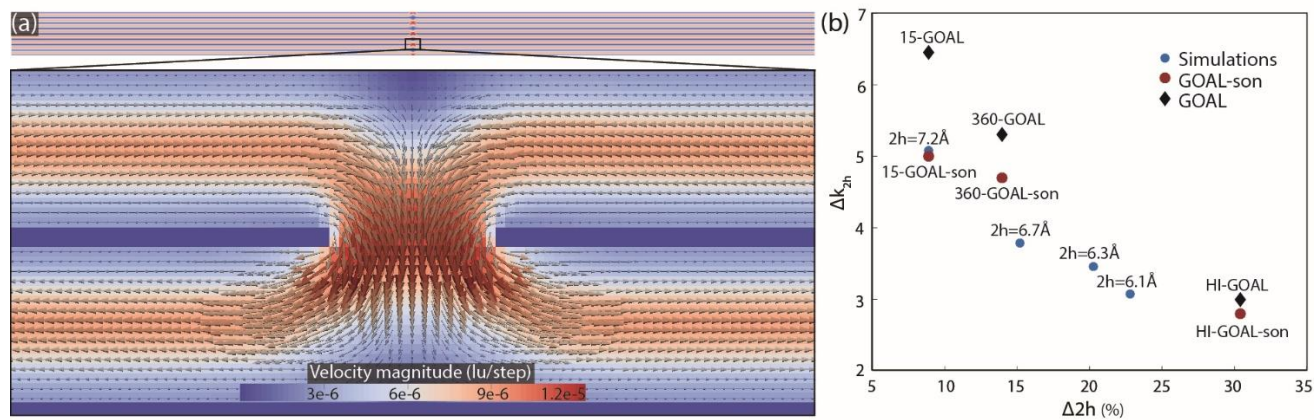

**Figure 5: LB simulations. (a)** GOAL geometry setup employed in the simulations and a magnification of the velocity field inside. **(b)** Normalized permeability vs interlayer spacing change for experimental and simulations outputs. The interlayer spacing change is normalized to the largest spacing value (7.9 Å).

In this letter, first we report on methods for independent fine tuning of GOAL dimensions (length and height of the GO nanochannels) via easy-to-implement techniques such as photo-reduction and bath sonication. The GO nanochannel height (i.e., GO spacing) was modified via UV irradiation in vacuo and HI treatment, which are more effective than the UV irradiation in ambient air. Second, the GO spacing was quantitatively correlated to the GO oxidation state allowing researchers to speed up the GO characterization process by performing a single technique to determine morphology and chemistry. Third, permeability tests highlight that GO interlayer spacing is the predominant dimension dictating GOAL water transport, although hydrophobicity effects may need to be taken into account. The experiments were corroborated with mesoscale simulations (LB approach), which represent a promising tool to evaluate water transport inside 2D materials. In summary, the results presented here offer more fundamental knowledge and thus the ability to rational design and engineer of GOAL membranes.

**Methods section**
**GOAL fabrication**
GO was synthesized by a modified Hummer's method and the detailed protocol can be found in our previous work.[38] Post synthesis, the GO solution was dispersed in ethanol at a concentration of 0.1 mg/mL. 1 mL of the GO solution was then dispersed in 20 mL of ethanol and vacuum filtered onto a PVDF membrane (Sterlitech). In the case of -GOAL-son, prior to VF, the GO solution was bath sonicated for 23 min in a Branson sonicator (V = 1.9 L, max power = 80 W, and f = 20 kHz). UV-GOAL was obtained by exposing the membrane to UV light (λ = 254 nm). The vacuum UV treatment was carried out in a TTP4-1.5K Probe Station (LakeShore) at $10^{-7}$ Torr. The HI treatment was performed by immersing the GOAL for 1 min in an aqueous concentrated HI solution (55%) at 50 °C. The membrane was then washed with 50 mL deionized water.

**Characterization**
**SEM.** The morphology of the bare PVDF, GO flakes, and GOAL was characterized using a Zeiss ULTRA Field Emission Scanning Electron Microscope with an In-lens secondary electron detector. The working distance was 4-5 mm and the acceleration voltage was 4 kV. The GO flake area was determined using ImageJ SEM analysis. In Figure 2, the *Adobe Photoshop* blend option was used at the border between bare PVDF and GOAL. **XPS.** GOAL chemistry was analyzed by a Thermo Scientific K-Alpha XPS (ESCA) with X-rays generated by a 12 kV electron beam with a spot size of 400 mm. The O/C ratio and peak deconvolution were quantified by Thermo Scientific Avantage software. The XPS instrumental error for atomic composition is ±1%, and the accuracy of the C1s peak fitting is ±2%. **AFM.** GOAL and bare PVDF morphology was characterized with an Asylum Cypher AFM using an Olympus 200TS cantilever (resonant frequency ≈ 150 kHz). The images were acquired in amplitude modulation mode[54] and the tip size was constantly monitored.[55] **XRD.** GOAL crystallographic structure was analyzed with a Bruker D8 equipped with a two-dimensional VANTEC-500 detector. The spectra were obtained by the integration of the 2D images (Figure S11) via EVA software. Depending on the sample, the integration time was between 600-1200 seconds. Note that the bare PVDF XRD spectrum is characterized by peaks that could overlap with GO. Thus, in order to increase signal-to-noise, the GOAL was cast and reduced on a porous alumina oxide ($d_{pore}$= 200 nm) from Whatman®. **SCA.** The contact angle measurements were completed with a Ramé-Hart 190 contact angle goniometer under ambient conditions. The SCA were evaluated with the Drop Analysis - DropSnake plugin in ImageJ. SCA were measured using 5 μL droplets and the data refer to the average of 5 measurements. **Raman.** Raman spectra were acquired using a WITec Confocal Raman Microscope/SNOM/AFM. The laser wavelength was 532 nm and the spectra are characterized using a 0.3 s integration time of at least 10 spectra.

**Simulations**
The numerical simulations are based on the Lattice Boltzmann (LB) method adapted using a novel Langevin-like frictional force to account for the GO-water H-bonding interactions. Since the LB method is largely documented in the literature,[56, 57] in the following we briefly introduce the LB model and the heterogeneous Langevin frictions term to account for the frictional interaction of hydroxyl and epoxy groups on water molecule as discussed in Eq. 3:

$$f_i(x + c_i \Delta t, t + \Delta t) = f_i(x,t) + \frac{1}{\tau}(f_i^{eq}(\rho, \rho u) - f_i(x,t)) + \Delta f \qquad \text{Eq. 3}$$

where $f_i$ is the probability distribution function at position *x* and time *t* with a lattice-constrained velocity $c_i$, where the index $^i$ runs over the nine directions of the lattice, $f_i^{eq}$ is a proper expansion of the Maxwell-Boltzmann distribution that can be expressed as a function of the density $\rho$ and linear momentum $\rho u$, the frictional force, $\Delta f$, has the following form:

$$\Delta f \approx -\rho \gamma(y) \boldsymbol{u} \qquad \text{Eq. 4}$$

where the gamma function $\gamma$ is as follows:

$$\gamma(y) \approx \gamma_0 \left(\exp\left(-\frac{y}{w}\right)\right) \qquad \text{Eq. 5}$$

where w = 0.2 nm is a representative size of the protruding functional groups and $\gamma_0$ is a characteristic water-hydroxyl collision frequency taken to be equal to 70 ps$^{-1}$.[58] Please refer to [59] for more details.

**Permeability**
The permeability tests were carried out in custom made dead end filtration system (Figure S7). The filtration system includes a 3.1 L reservoir (e.g., pressure pot from Alloy Product Corp.) allowing for large volume experiments to be performed. The pressure was regulated with a pressure gauge (Ingersoll). The pure water permeability was evaluated by monitoring the permeate flux with a Sartorious Laboratory balance every 30 min. The was pure water permeability was then obtained by dividing the flow rate by the applied pressure.

**Acknowledgments**
This work made use of the Center for Nanoscale Systems at Harvard University, a member of the National Nanotechnology Infrastructure Network, supported (in part) by the National Science Foundation under NSF award number ECS-0335765
.

**Supporting Information**
The SI includes further GOAL characterization (AFM, SEM, Raman, and XPS), results of the GOAL UV reduction in air, photographs of the GOAL membrane and permeability setup, and tables summarizing the GOAL permeability changes due to chemical reduction and/or sonication. The Supporting Information is available free of charge via the Internet at http://pubs.acs.org.

**ELECTRONIC SUPPORTING INFORMATION**

Figure S1 represents AFM image of the GOAL membranes. In particular, the images are taken at the boundary (highlighted with a black line) between the GO and the bare PVDF. From the topography image (Fig S1a) it is possible to notice that the thin GO layer preserve the flower-like structure of the PVDF membrane. Important information can also be derived from the AFM phase image (Fig S1b). It is well know that the phase signal [54] is affected by the chemistry of the sample, thus, heterogeneity in the chemistry can be recognized. In our case, GO can be easily reciognized due to its different chemical composition compared ot the bare PVDF.

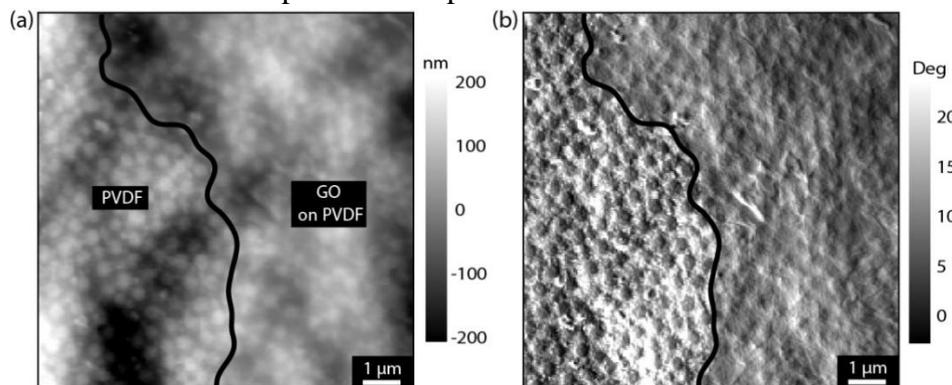

**Figure S 1: GOAL characterization by AFM. (a)** topography and **(b)** phase image acquired at the boundary between the GO layer and the bare PVDF via AFM. The boundary between the two regions is highlighted with a continuous black line.

.

Figure S2 represents a photo of the GOAL membranes treated with different reduction techniques. The GOAL membranes are characterized by a diameter of 2.2 cm and it is possible to see the effect of the reduction which recovers the graphitic nature of the membranes.

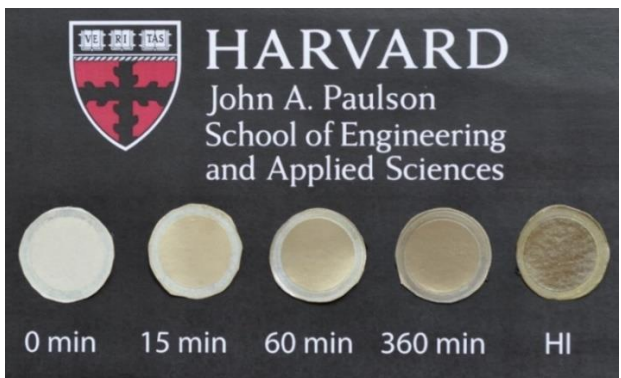

**Figure S 2: GOAL membranes treated with different reduction techniques.**

Table S1 reports the oxygen and carbon footprints for the GOAL membranes. It is important to underline that the HI treatment is the most effective reduction technique among the ones tested and it allows achieving O/C ratio of 21%. Regarding the UV irradiation, longer irradiation times lead to stronger reduction and the vacuum dictates more effective reduction compared to air.

Table S1: reduction mechanisms. The O/C ratio values are characterized by instrumental error of ±1%, whereas the C1s peak deconvolution is affected by an instrumental error of ±2%.

| Treatment | Time (min) | Epoxide/hydroxide | Carbon | Carboxylate | O/C |
|---|---|---|---|---|---|
| Untreated | 0 | 68.56 | 27.93 | 3.51 | 0.71 |
| UV vacuum | 20 | 57.27 | 37.16 | 5.57 | 0.58 |
|  | 60 | 34.12 | 57.81 | 8.07 | 0.52 |
|  | 360 | 26.73 | 61.91 | 11.36 | 0.46 |
|  | 720 | 21.75 | 70.62 | 8.03 | 0.39 |
|  | 1440 | 13.87 | 73.48 | 12.65 | 0.35 |
| UV air | 60 | 37.79 | 51.5 | 10.67 | 0.65 |
|  | 360 | 38.12 | 51.05 | 10.83 | 0.56 |
|  | 1440 | 19.28 | 66.21 | 14.51 | 0.46 |
| HI | 1 | 11.69 | 76.46 | 11.85 | 0.21 |

Figure S3a displays the evolution of the C1s spectrum with UV irradiation time in air. As explained in the main text, longer irradiation time leads to a more efficient graphitization of the GOAL. The intensity of the single (C-C) and double carbon (C=C) bond binding increases from 28% to 51%. This is also confirmed by the oxygen to carbon mass ratio (i.e., O/C) which decreases from 72% to 56% for untreated GOAL and the UV-GOA irradiated for 360 min in air, respectively. Figure S3b represents the evolution of the XRD signal of GOAL with different UV irradiation times. We observed that the decrease in the number of functional groups with an increase in the exposure time leads to a shift of the peak of circa 1 Å. However, as explained in the main text, the air treatment leads to a nosier XRD signal.

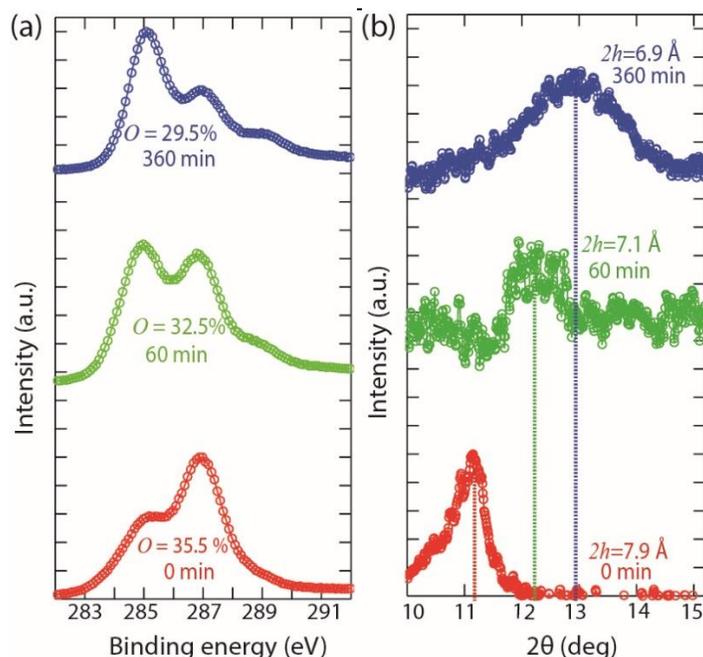

**Figure S 3: GOAL oxygen content and interlayer spacing via UV reduction in air.** (**a**) XPS and (**b**) XRD spectra evolution for the UV irradiated GOAL membranes in air.

Figure S4 represents the XPS and XRD spectra for HI-GOAL. We did not succeed in reducing the GO on $Al_2O_3$ with HI and for this reason we used PVDF as the substrate. As explained in the main text, the PVDF exhibits peaks in the same region of GO. However, in Figure S4a it is possible to notice the appearance of a peak centered at circa 16º. This peak, which was not as strong in the bare PVDF, can be related to the presence of the HI-GOAL and corresponds to a spacing of 5.5 Å.

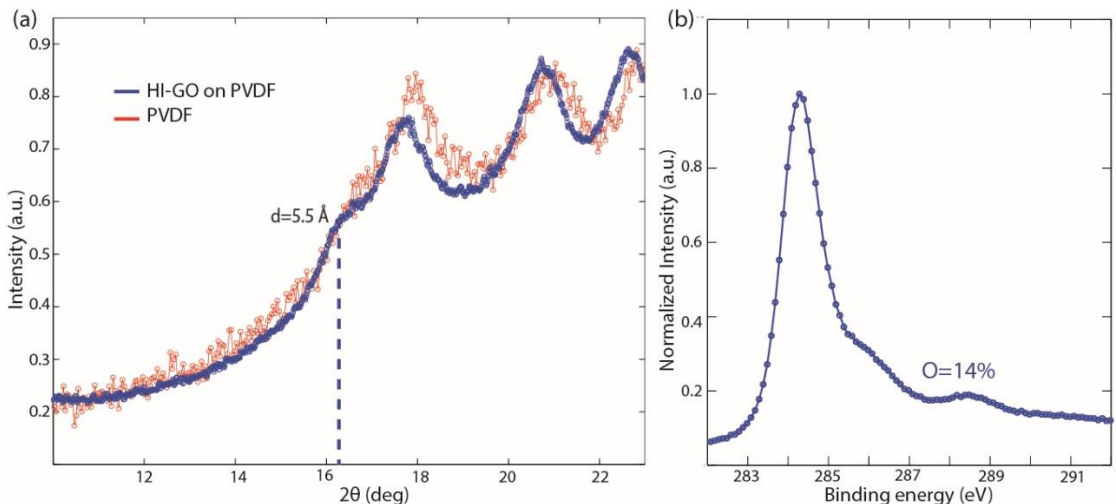

**Figure S 4: GOAL oxygen content and interlayer spacing via HI treatment. (a)** XRD and **(b)** XPS spectrum for the GOAL treated with HI.

Figure S5 represents the effect of the sonication treatment on the GO flake size. In particular, with 23 min of bath sonication, the flake size varies from (52.2 ±18.9) µm² to (1.3 ±0.4) µm².

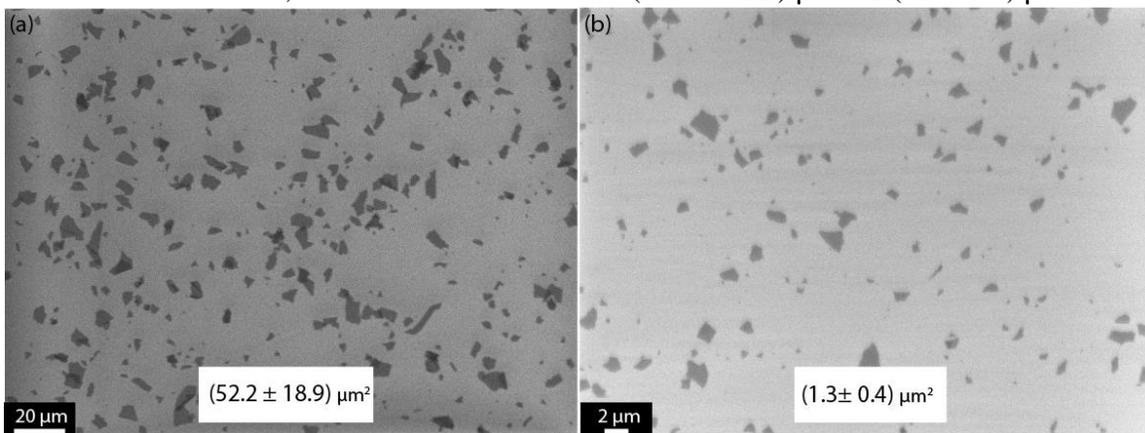

**Figure S 5: SEM analysis of the sonication treatment.** GO flakes size **(a)** before and after **(b)** ultrasonic irradiation.

Figure S6 represents the independence between the variation of the dimensions of the GOAL architecture. Figure S6a displays the variation of the GO flake size versus the UV treatment time in vacuum. Although the UV treatment is responsible for the changing of the GOAL chemistry (i.e., the GOAL spacing), it is possible to notice that even 360 min of UV irradiation does not lead to a significant variation of the GO flake size. At the same time, the bath sonication treatment, responsible for the changing of the GO flake size, does not significantly change the chemical composition of the GOAL (Figure S6b).

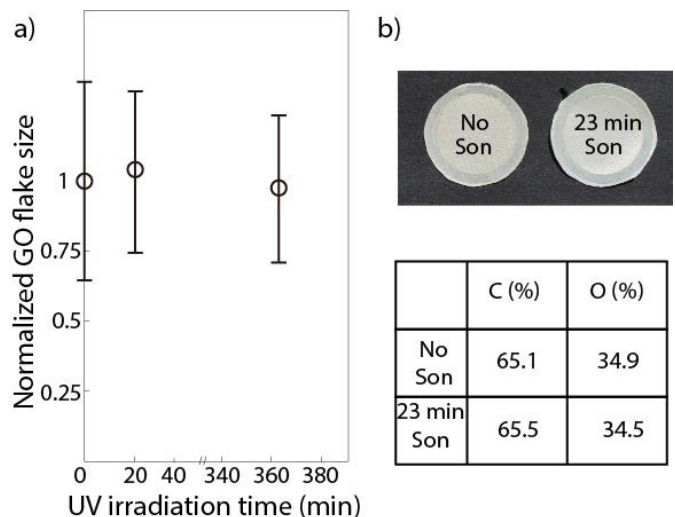

**Figure S 6: Absence of interdependency between the tuning of flake size and oxygen content. a)** variation of the GO flake size with UV irradiation in vacuum. **(b)** Photo of GOAL obtained with sonicated and not-sonicated GO solution and their atomic composition**.**

Figure S7 illustrates the dead-end filtration system, which is characterized by a reservoir (i.e., pressurized pot), allowing the filtration experiments to be performed with large volume (i.e., more than 3 L). The pressure taken from the wall was regulated with a pressure gauge form *Ingersoll*. The membranes were placed in a stainless steel EDM Millipore holder.

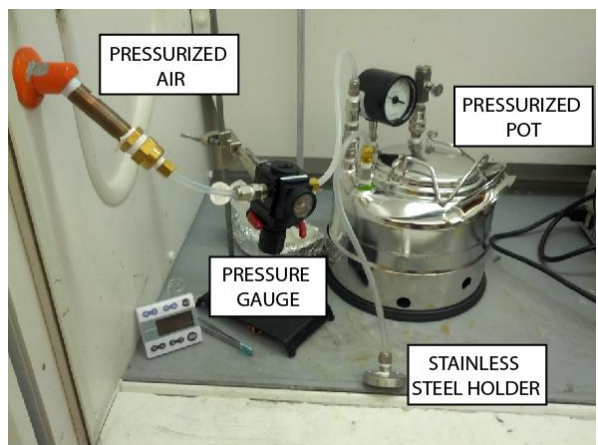

**Figure S 7: Dead-end filtration set-up.**

Figure S8 represents the water permeation through a 0-GOAL membrane versus time. The initial permeability is characterized by higher values compared to the permeability at steady state conditions, which are reached after circa 700 min of filtration. In particular, during the membrane compaction stage, the permeability reduces by circa three-fold. The Darcy behavior of the membrane can also be observed with a pressure decrease from 50 to 20 psi, which leads to a reduction of permeability of almost a factor of three.

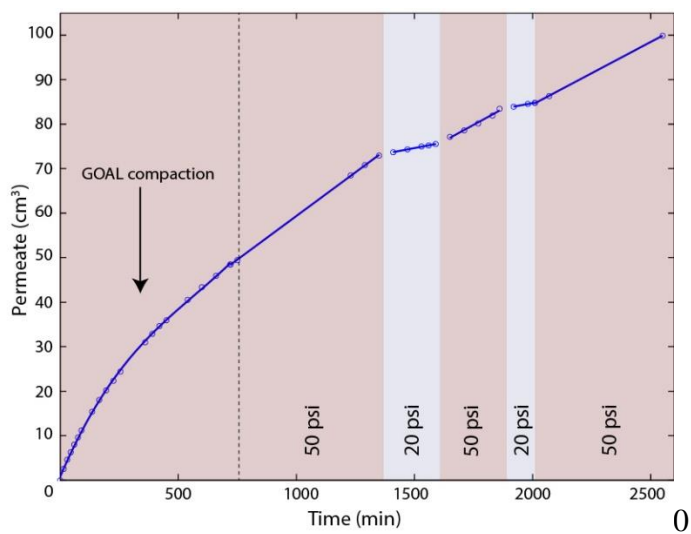

Figure S 8: Water permeation through 0-GOAL versus time.

In Figure S9, the GOAL structure was characterized via Raman spectroscopy and display a D-band (A1g symmetry) at ~1350 cm$^{-1}$ representative of defects/disorder in the basal plane and a G-band (E2g symmetry) at ~1590 cm$^{-1}$ corresponding to the in-plane sp$^2$ bond stretching, thus proportional to the extension of the graphitic domains. The longer exposure to UV irradiation leads to a smaller D/G peak intensity ratio ($I_D/I_G$). In particular, the $I_D/I_G$ is equal to 0.95 and 0.90 for 15 min and 720 min UV irradiation, respectively, confirming the restoring of the graphitic domains via the reduction of GOAL.

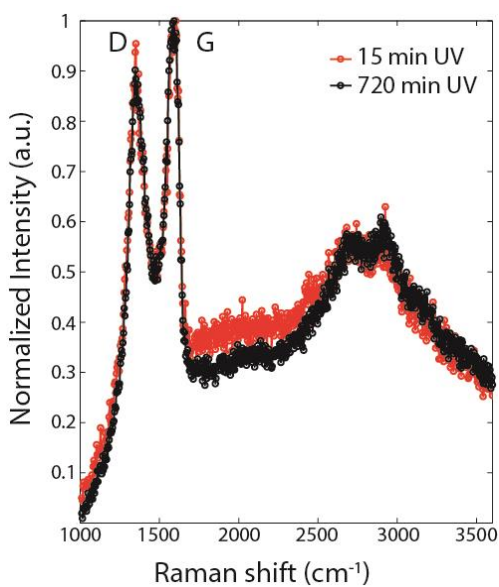

Figure S 9: Raman spectra of the D- and G-bands for 15minUV-GOAL and 720minUV-GOAL.

Table S2 and S3 represent the variation in the permeability ($\Delta_k$) due to the chemical reduction and sonication, respectively. The tables also report the normalized variations ($\Delta_{2h}$ and $\Delta_l$) for each scenario. As explained in the main text, the larger values of $\Delta_{2h}$ highlight the importance of the nanochannels interlayer spacing in dictating the permeability, compared to the length of the nanochannels.

Table S2: effect of the GO nanochannels' interlayer spacing (*2h*) on the permeability.

|  | 15-GOAL | 360-GOAL | HI-GOAL | 15-GOAL-son | 360-GOAL-son | HI-GOAL-son |
|---|---|---|---|---|---|---|
| $\Delta_k$ (%) | 56 | 74 | 91 | 43 | 66 | 86 |
| $\Delta k_{2h}$ | 6.35 | 5.3 | 3.0 | 5.0 | 4.7 | 2.8 |

Table S3: effect of the GO nanochannels' length (*l*) on the permeability.

|  | 0-GOAL-son | 15-GOAL-son | 360-GOAL-son | HI-GOAL-son |
|---|---|---|---|---|
| $\Delta_k$ (%) | 19 | 53 | 60 | 110 |
| $\Delta k_l$ | 0.02 | 0.08 | 0.09 | 0.15 |

Eq. S1 and Figure S10 display the exponential relationship between GOAL SCA and the degree of GO oxidation:

$$SCA = a * \text{Exp}(-b * O) + c \qquad \text{Eq. S1}$$

where *a*, *b,* and *c* are empirical coefficients with values of -2.57, -3.47, and 82.6, respectively.

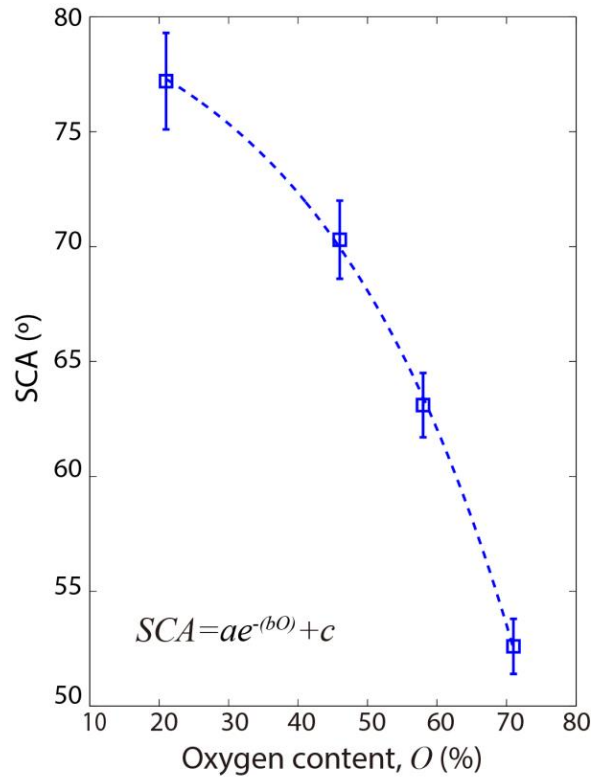

**Figure S10: GOAL SCA versus oxygen content.** In the equation, a, b, and c are empirical coefficients with values of -2.57, -3.47, and 82.6, respectively.

Figure S11 represents the 2D XRD images on which the results in Figure 3b are based. From the integration of these images the XRD spectra can be obtained.

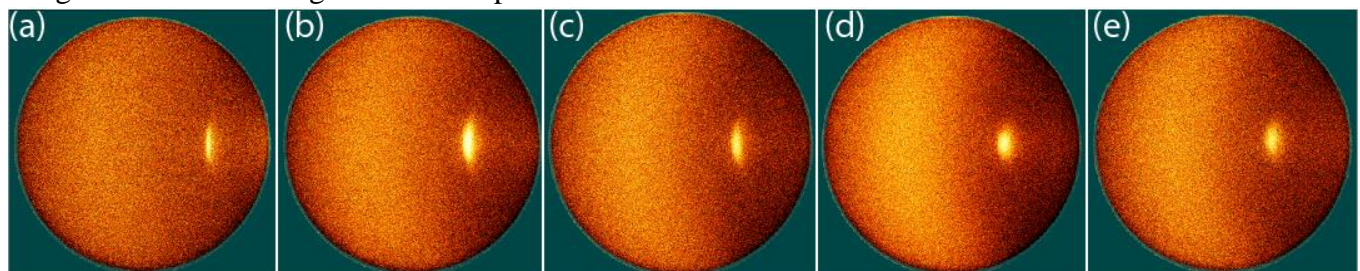

**Figure S11: 2D XRD images (a-e)** Images for 0min,15min, 60min, 360min and720 min UV irradiatied-GOAL in vacuum.